\documentclass[aps,prl,twocolumn]{revtex4}
\usepackage{epsfig}

\begin{document}

\title{Mapping spin coherence of a single rare-earth ion in a crystal onto a single photon polarization state}

\author{Roman Kolesov, Kangwei Xia, Rolf Reuter, Rainer St\"ohr, Tugrul Inal, Petr Siyushev, and J\"org Wrachtrup}

\address{3. Physikalisches Institut, Universit\"at Stuttgart and 
Stuttgart Research Center of Photonic Engineering (SCoPE), 
Pfaffenwaldring 57, Stuttgart, D-70569, Germany
}

\begin{abstract}
We report on optical detection of a single photostable $Ce^{3+}$ ion in an yttrium aluminium garnet (YAG) crystal and on its magneto-optical properties at room temperature. The quantum state of an electron spin of the emitting level of cerium ion in YAG can be initialized by circularly polarized laser pulse. Furthermore, its quantum state can be read out by observing temporal behaviour of circularly polarized fluorescence of the ion. This implies direct mapping of the spin quantum state of $Ce^{3+}$ ion onto the polarization state of the emitted photon and represents one-way quantum interface between a single spin and a single photon.
\end{abstract}

\maketitle

Rare-earth doped optical materials are known to have outstanding properties for optical information storage and processing, both classical \cite{classical_storage} and quantum \cite{PrYSO_long_storage,single_photon_storage,entanglement_storage}. In that respect, the most valuable feature of rare-earth ions is that the quantum transitions between their electronic states and nuclear hyperfine levels have extremely high quality factor due to efficient screening of their optically active $4f$ electrons from the surrounding crystalline environment by outer lying $5s$ and $5p$ electronic shells. As a result, rare-earth ions in optical crystals have demonstrated benchmark performance in storing quantum information for over a second \cite{PrYSO_long_storage}, storing and retrieving quantum state of a single photon \cite{single_photon_storage}, and quantum entanglement \cite{entanglement_storage}, etc. In these applications, the quantum state of a single flying qubit (a single photon) or of a pair of entangled flying qubits (a pair of entangled photons) is mapped onto an ensamble of rare-earth ions. However, another challenging task of interfacing a single flying qubit with a single stationary qubit is very hard to address. The reason is that the high Q-factor of electronic transitions of rare-earth ions inevitably results in very low photon flux emitted by an individual ion. This, in turn, makes single rare-earth ions in crystals very hard to detect. A solution to this problem \cite{Single_Pr} is to detect the emission originating from strong parity-allowed $4f^{(n-1)}5d\rightarrow 4f^n$ transition yielding high photon flux.

In the present work, we optically address for the first time single electron spins of $Ce^{3+}$ ions in a crystal. We show that the electron spin of an ion can be prepared in a coherent state by circularly polarized laser excitation. Furthermore, it is shown that spin coherence can be efficiently mapped onto the polarization state of a photon emitted by a $Ce^{3+}$ ion. Thus, we demonstrate a one-way coherent interface between a single spin and a single photon.

$Ce^{3+}:YAG$ is a very efficient scintillator emitting in green-yellow spectral range (peak emission is at $550\;nm$) with quite short lifetime ($\sim 60\;ns$). The fluorescence of $Ce^{3+}$ can be efficiently excited by optical pumping into the phonon absorption sideband of the lowest $5d$ level with the peak absorption around $460\;nm$. Taking into account $60\;ns$ lifetime of the fluorescence and almost unity quantum efficiency \cite{CeYAG_quantum_efficiency}, a single $Ce^{3+}$ ion is expected to emit $1.6\times 10^7$ photons per second. This amount of light can be easily detected by means of confocal microscopy. In order to distinguish individual ions in a crystal, the distance between them should be greater than the resolution of the microscope resulting in a maximum concentration of cerium of $36\;ppb$ (parts per billion) relative to yttrium. This value sets the absolute upper limit on the concentration of cerium impurity. In turn, any yttrium-based crystal contains trace amounts of all rare-earth elements including cerium, so that only the YAG crystals of highest purity can be used for isolation of a single emitting ion.

\begin{figure}
\center{
\includegraphics[width=8cm]{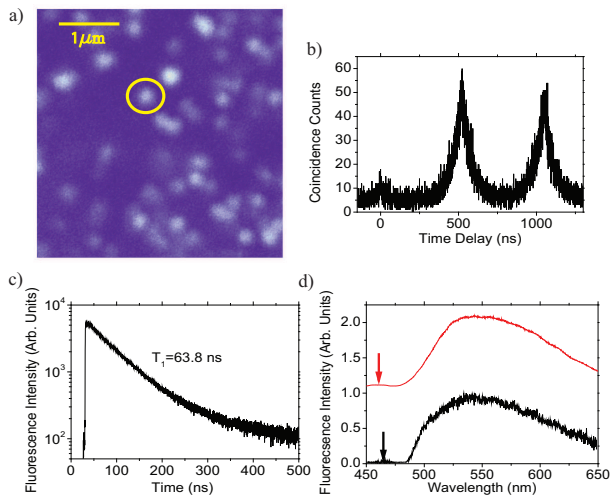}
\caption{\label{fig:scan} Detection of a single $Ce^{3+}$ ion in a YAG crystal. a) Scanning confocal image of $Ce^{3+}$ centers in a nominally pure YAG crystal. b) Photon correlation signal taken on the marked spot on Figure a) with a pulsed laser excitation. The peak at zero time delay is much weaker than the other two indicating that the emitter is indeed a single quantum object. c) Fluorescence decay signal taken on the same spot reveals the lifetime of $64\;ns$ in agreement with the known lifetime of lowest $5d$ state of $Ce^{3+}$ in YAG. d) The spectrum of the emission of the same spot (lower curve, black line) well correlates with the fluorescence spectrum of a bulk $Ce:YAG$ crystal (upper curve, red line). The arrows indicate the excitation wavelength.}
}
\end{figure}

In the present study, we used the same crystal in which we detected single praseodymium ions in our previous publication \cite{Single_Pr} (ultra-pure YAG crystal produced by Scientific Materials Corp., boule 14-12). Optical detection of a single cerium center was performed in a home-built confocal microscope operating at room temperature (see Supplementary Fig. S1). We used frequency-doubled output of a femtosecond Ti:Sapphire laser operating at $\approx 920\;nm$ as an excitation source. The repetition rate of frequency-doubled pulses was reduced from the original $76\;MHz$ to a few $MHz$ rate by a pulse-picker. The fluorescence of cerium ions was split into two paths and detected by two single-photon-counting avalanche photodiodes. In this way, a Hanbury-Brown and Twiss setup was arranged in order to allow photon correlation measurements. The detection wavelength range was restricted between $485\;nm$ and $630\;nm$. While the short wavelength boundary is needed to reject the excitation light, the long wavelength cut-off was introduced in order to block fluorescence of $Cr^{3+}$ impurity ions present in the crystal. The resulting scanning image of the crystal is shown in Fig.\ref{fig:scan}a. Each individual bright spot corresponds to a single cerium ion. Several tests were performed to confirm that: 1) photon corellation measurements performed on these spots show pronounced antibunching at zero delay time between the count on two detectors indicating a single quantum emitter giving rise to the fluorescence (see Fig.\ref{fig:scan}b); 2) the fluorescence lifetime of each spot is close to $60\;ns$ in agreement with the lifetime data taken on highly doped $Ce:YAG$ crystals (see Fig.\ref{fig:scan}c); 2) the emission spectrum is the same as that of the  highly doped $Ce:YAG$ crystals (see Fig.\ref{fig:scan}d). The typical photon count rate on the two detectors produced by a single $Ce^{3+}$ ion was $40-50\; kcounts/s$ with the pulse repetition rate $7.6\;MHz$. The ions are photostable for many hours of continuous illumination and show no blinking.

\begin{figure}
\center{
\includegraphics[width=8cm]{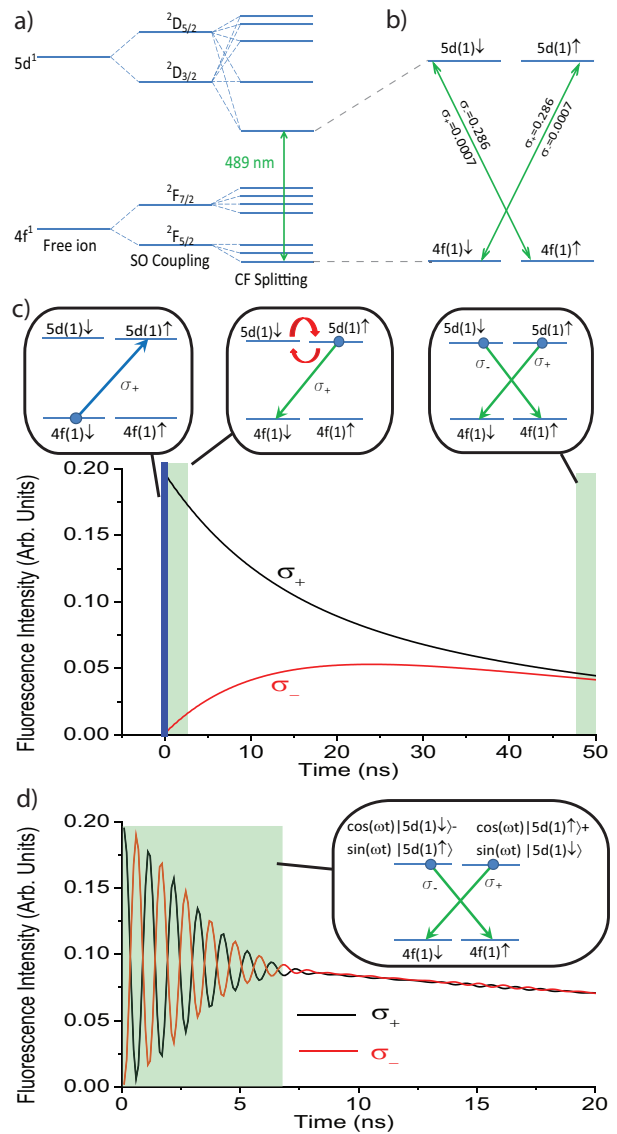}
\caption{\label{fig:levels1} Electronic level structure of $Ce^{3+}$ ion in a YAG crystal. a) Electronic $4f^1$ and $5d^1$ shells are split by a combined action of spin-orbit coupling and the crystal field. While in the lower $4f^1$ state spin-orbit coupling dominates, in the upper $5d^1$ states the major contribution to the splitting is from the crystal field. It causes significant red shift of the positions of the $5d$ levels. b) Strengths of the transition dipoles between the lowest $4f$ and the lowest $5d$ Kramer's doublets. The quantization axis is chosen along the z-axis of the local frame of the cerium site \cite{YAG_orientation}. c) Calculated time traces of the fluorescence corresponding to the ZPL of $Ce^{3+}$ once the ion is pumped into one of the lowest $5d$ spin sublevels (left inset). The magnetic field is along the z-axis of the local frame of cerium site. The $\sigma_-$ polarization (red curve) builds up due to spin relaxation (central inset) while the $\sigma_+$ polarization (black curve) declines until it reaches the level of $\sigma_-$ polarization (right inset). d) Calculated fluorescence oscillations with the external magnetic field applied along the x-axis of the local frame of cerium site. Randomly oriented hyperfine field produced by the $^{27}Al$ nuclei causes fast dephasing.}
}
\end{figure}

In order to get insight into the optically excited spin dynamics of cerium in YAG, we consider the electronic energy level structure of $Ce^{3+}:YAG$ in detail. $Ce^{3+}$ ion has only one unpared electron whose ground state is $4f^1$. The 14-fold degeneracy of that state (7-fold degeneracy due to the orbital momentum times doubly degenerate spin) is lifted by the combined action of spin-orbit coupling and the crystal field (see Fig.\ref{fig:levels1}a). Overall, the 14-fold degenerate $4f^1$ manifold is split into 7 spin doublets (so called Kramer's doublets) which are grouped into two sub-manifolds, $^2F_{5/2}$ and $^2F_{7/2}$, containing 3 and 4 doublets respectively. The energy difference between $^2F_{5/2}$ and $^2F_{7/2}$ is mostly defined by spin-orbit coupling while the splitting within the manifolds is determined by the crystal field. The degeneracy of the Kramer's doublets can only be lifted by an external magnetic field. On the contratry, the structure of the excited $5d$ electronic levels is dominated by crystal field splitting slightly perturbed by spin-orbit interaction. For the $5d$ levels the electron spin thus is a good quantum number while in the $4f$ states the spin is highly mixed with the orbital momentum. This, in turn, suggests that the optical transitions between the ground $4f$ and the excited $5d$ levels should be accompanied by a flip of a $Ce^{3+}$ electron spin. Owing to its spin-flip optical transitions, $Ce^{3+}:YAG$ crystals exhibit strong Faraday effect, i.e. rotation of polarization of the optical field propagating through the crystal with the external magnetic field applied \cite{CeYAG_Faraday}. The inverse effect, i.e. magnetizing the material by circularly polarized optical field, was demonstrated experimentally previously by direct measurement of the magnetic moment of $Ce:YAG$ crystal induced by a circularly polarized laser pulse \cite{CeYAG_magnetization}. In this work, the magnetization was attributed to non-equilibrium population of the two spin levels of $Ce^{3+}$ ion in its lowest $5d$ state. Furthermore, it was shown that the non-equilibrium spins experiences coherent Larmor precession in the externally applied magnetic field. However, direct detection of the magnetization can only be used to study the coherent spin behaviour of an ensamble of spins, but inapplicable to study an individual cerium ion since its magnetic signal is too weak to be detectable. On the other hand, since the $Ce^{3+}$ spin can be oriented by chosing the appropriate polarization of the excitation light, the polarization of the cerium ion fluorescence also depends on the spin state of the emitting $5d$ level. A quantitative understanding of this dependence requires the knowledge of the orientations of the dipoles associated with the optical transitions between the $4f$ and the lowest $5d$ Kramer's doublets.The result of a detailed calculation (see Supplementary Material) is depicted in Fig.\ref{fig:levels1}b and in Supplementary Fig. S2. Since the ratio of the oscillator strengths of the transitions $\left|4f(1)\downarrow\right\rangle\leftrightarrow \left|5d(1)\uparrow\right\rangle$ and $\left|4f(1)\uparrow\right\rangle\leftrightarrow \left|5d(1)\downarrow\right\rangle$ is $0.284/0.0007\approx 400$ for $\sigma_+$/$\sigma_-$ polarizations, circularly polarized light can excites $Ce^{3+}$ ion from its lowest $4f$ doublet into one of the $5d$ spin states 400 times more efficiently than into the other spin state resulting in almost perfect spin polarization in the excited state. In turn, the emission originating from one of the spin sublevels of the lowest $5d$ state and terminating at the lowest $4f$ doublet should be circularly polarized. This transition corresponds to the zero-phonon line (ZPL) of $Ce^{3+}:YAG$ at $489\;nm$ \cite{CeYAG_ZPL}. Any relaxation process leading to spin flip of the $5d$ emitting state should give rise to the opposite circular polarization of the emitted fluorescence as shown in Fig.\ref{fig:levels1}c.  If an external magnetic field is applied perpendicular to the system quantization axis, the excited state spin starts precessing at Larmor frequency defined by the magnitude of the field and the g-factor of the $5d(1)$ state: $\omega=\mu_B g_{\perp} B_{\perp}/\hbar$.
This leads to oscillatory behaviour of the $\sigma_+$ and $\sigma_-$ components of the fluorescence (see blue curve on Fig.\ref{fig:levels1}c). Any decoherence process shortens the lifetime of oscillations (see Fig.\ref{fig:levels1}d). In case of $Ce^{3+}$ ion in YAG the dephasing arises from the hyperfine interaction with the surrounding $^{27}Al$ nuclei producing random magnetic field at the location of the cerium ion. This magnetic field is estimated to be $\approx 40\;G$ \cite{CeYAG_magnetization}.

\begin{figure}
\center{
\includegraphics[width=8cm]{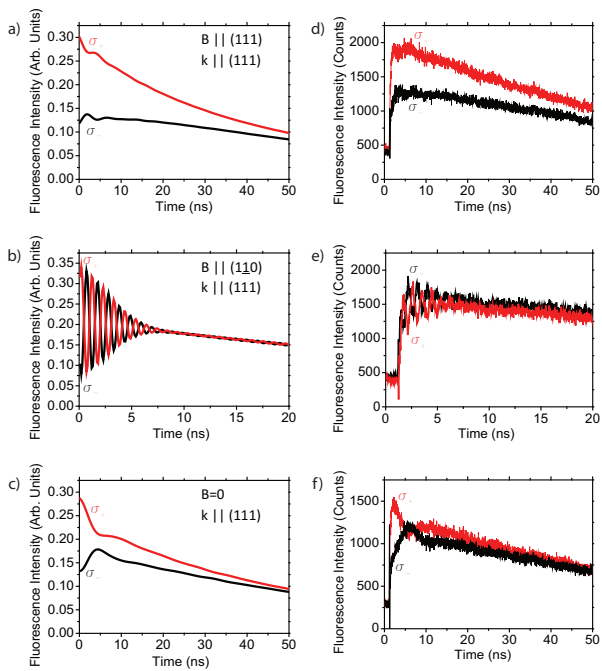}
\caption{\label{fig:beats1} Quantum beats of the fluorescence of a single $Ce^{3+}$ ion. a)-c) Theoretically calculated fluorescence decay traces for $\sigma_+$- and $\sigma_-$-polarized emissions. The excitation and detection are assumed to be along the (111) axis of the YAG crystal. The emission is assumed to terminate at either of the three $^2F_{5/2}$ Kramer's doublets. a) The magnetic field of $70\;G$ is along the (111) axis. b) The magnetic field of $400\;G$ is along the (1\=10) crystal axis. c) No external magnetic field is applied. The non-exponential behaviour of the fluorescence decay is due to the hyperfine coupling with the surrounding aluminium nuclei. d)-f) Experimentally measured fluorescence decay curves. The experimental conditions are identical to the ones assumed in a)-c).}
}
\end{figure}

Unfortunately, the ZPL of $Ce^{3+}:YAG$ is not observable at room temperature, i.e. the emission into the lowest $4f$ Kramer's doublet (see Fig.\ref{fig:levels1}b) cannot by discriminated from the emission terminating at other $4f$ states. Indeed, the presence of the other $4f$ states deteriorates the contrast of the spin-dependent fluorescent signal. This can be seen from the fact that the sum of the dipole strengths corresponding to the emission of the two circularly polarized components are equal for both $5d$ spin states (see Supplementary Table S1). Thus, by detecting the full fluorescence signal it is impossible to reveal the emitting spin state of the $5d$ level. However, spectral filtering of the emission improves the situation. In particular, the emission terminating at the levels belonging to the $^2F_{5/2}$ manifold can be separated by detecting only the wavelengths shorter than $550\;nm$. Under these circumstances, the contrast of the spin dependence of the circularly polarized fluorescence signal can be restored. The calculated time traces corresponding to the experimental situation are indicated in Figs.\ref{fig:beats1}a-c. Here it is assumed that the excitation of cerium ions and the detection of their fluorescence are arranged along the (111) crystallographic axis of YAG ((111) was the orientation of the crystal used in the experiment) while the external magnetic field is applied either parallel or perpendicular to this axis. The spin relaxation at the rate $1/28\;ns^{-1}$ \cite{CeYAG_magnetization} and spin dephasing in the random hyperfine magnetic field are taken into account.

Experimental observation of the excited state dynamics of cerium ions is now straightforward. Individual ions identified in the confocal microscope are now excited with a circularly polarized laser pulses. The polarization of the fluorescence detected within the window $485\; nm$ through $550\; nm$ is also chosen to be either $\sigma_+$ or $\sigma_-$. An external magnetic field is applied either parallel or perpendicular to the excitation beam propagation direction. The results of measurements are shown in Fig.\ref{fig:beats1}d-f. With the magnetic field oriented along the propagation direction of the excitation beam, the amounts of the fluorescence corresponding to $\sigma_+$ and $\sigma_-$ polarizations significantly differ in the beginning of the decay indicating non-equilibrium populations of the excited spin-up and spin-down states. The decay traces merge after some time. This means that the spin populations equilibriate due to some spin-flip processes, being most probably related to Orbach or phonon Raman relaxation. The exponential fit to the difference between the spin-up and spin-down decay signals gives the decay constant of $T_1=28\;ns$. This spin-lattice relaxation lifetime is with excellent agreement with the previously reported value \cite{CeYAG_magnetization}. In case of the magnetic field being perpendicular to the excitation beam propagation direction, one sees rapid oscillations of the fluorescence reflecting Larmor precession of the electron spin. The decay curves originating from spin-up and spin-down $5d$ states and corresponding to the $\sigma_+$ and $\sigma_-$ emissions, respectively, show out-of-phase oscillations, as expected. When no external magnetic field is applied, the decay signals corresponding to $\sigma_+$ and $\sigma_-$ polarizations show non-exponential behaviour due to the hyperfine interaction of cerium spin with aluminium nuclei (see Fig.\ref{fig:beats1}f).

In summary, we have obtained detection of individual $Ce^{3+}$ ions in a crystal confirmed by spectral, lifetime, and photon anti-bunching measurements. Most importantly, single cerium ions can be prepared in a well-defined spin state of the lowest $5d$ level by optical pumping with circularly polarized light and the coherent state of the spin can be read out by observing the dynamics of the emitted fluorescence. The latter fact opens a way to transfer coherent state of an electron onto the polarization of an emitted photon. The mentioned results were obtainedat room temperature at which $Ce^{3+}$ spin relaxes fast in the ground state due to efficient Orbach process \cite{Orbach}. Cooling the crystal down to cryogenic temperatures would open a possibility of all-optical addressing of cerium spin in the ground $4f$ state thus allowing for single spin optical memory similar to the one previously demonstarted using quantum dots \cite{single_spin_memory}. In turn, optical detection of a single $Ce^{3+}$ ion and study of its magneto-optical properties in low-spin crystal hosts like, for example, yttrium orthosilicate \cite{CeYSO} would dramatically increase the decoherence time. Furthermore, addressing the ion emission into its ZPL would allow one to realize spin-photon entanglement \cite{spin-photon_entanglement} in rare-earth-doped crystalline material.

\end{document}